\documentclass[10pt,aps,twocolumn,prc,superscriptaddress,showpacs,nofootinbib,noshowkeys,longbibliography,preprintnumbers]{revtex4-1}
\usepackage[dvips]{graphics,graphicx}
\usepackage{color}
\usepackage{amsmath, amssymb}
\usepackage{multirow}
\usepackage{longtable}
\usepackage[normalem]{ulem}  

\def\lsim{\raise0.3ex\hbox{$<$\kern-0.75em\raise-1.1ex\hbox{$\sim$}}}
\def\gsim{\raise0.3ex\hbox{$>$\kern-0.75em\raise-1.1ex\hbox{$\sim$}}}
\begin{document}
\title{Effects of  kinematic  cuts on net-electric charge fluctuations}
\date{\today}
\author{Frithjof Karsch}
\email{karsch@bnl.gov}
\affiliation{Fakult\"at f\"ur Physik, Universit\"at Bielefeld, D-33615 Bielefeld,
Germany}
\affiliation{Physics Department, Brookhaven National Laboratory, Upton, New York 11973, USA}
\affiliation{Key Laboratory of Quark \& Lepton Physics (MOE) and Institute
of Particle Physics, Central China Normal University, Wuhan 430079, China}
\author{Kenji Morita}
\email{kmorita@yukawa.kyoto-u.ac.jp}
\affiliation{Yukawa Institute for Theoretical Physics, Kyoto University,
Kyoto 606-8502, Japan}
\affiliation{Institute of Theoretical Physics, University of Wroclaw,
PL-50204 Wroc\l aw, Poland}
\author{Krzysztof Redlich}
\email{krzysztof.redlich@ift.uni.wroc.pl}
\affiliation{Institute of Theoretical Physics, University of Wroclaw,
PL-50204 Wroc\l aw, Poland}
\affiliation{ExtreMe Matter Institute (EMMI), 64291 Darmstadt, Germany}
\preprint{YITP-15-60}
\affiliation{%
Department of Physics, Duke University, Durham,  North Carolina 27708, USA}
\begin{abstract}
 The effects of kinematic cuts on electric charge fluctuations in a gas
 of charged particles are discussed.  We consider a very transparent
 example of an ideal pion gas with quantum statistics, which can be
 viewed as a multi-component  gas of Boltzmann particles with
 different charges, masses, and degeneracies. Cumulants  of
 net-electric charge fluctuations $\chi_n^Q$ are calculated  in a static and
 expanding medium with flow parameters adjusted to the experimental data. We
 show that the transverse momentum cut, $p_{t_\text{min}}\leq p_t\leq
 p_{t_\text{max}}$,  weakens the effects of Bose statistics, i.e.
 contributions of effectively multi-charged states to higher order moments.
 Consequently, cuts in $p_t$ modify the experimentally measured cumulants and
 their ratios. We discuss the influence  of kinematic cuts on the
 ratio of mean and variance of electric charge fluctuations in a hadron
 resonance gas, in the light of recent data of the STAR and PHENIX
 Collaborations. We find that the different momentum cuts of
 $p_{t_\text{min}}=0.2$~GeV (STAR) and $p_{t_\text{min}}=0.3$~GeV
 (PHENIX) are   responsible for more than 30\% of the difference between
 these two data sets. We argue that the $p_t$ cuts imposed on charged
 particles will influence the normalized kurtosis
 {$\kappa_Q\sigma_Q^2 = \chi_4^Q/\chi_2^Q$} of the electric charge fluctuations.
 In particular, the reduction of $\kappa_Q\sigma_Q^2$ with increasing
 $p_{t_\text{min}}$ will lead to differences between  PHENIX and STAR data
 of ${\cal O}(6\%)$ which currently are buried under large
 statistical and systematic errors. We furthermore introduce the
 relation between momentum cut-off and finite volume effects, which is
 of relevance for the comparison between experimental data and lattice
 QCD calculations.
\end{abstract}
\maketitle


\section{Introduction}
Fluctuations of conserved charges provide a unique opportunity to describe
the state of matter created in heavy ion collisions
\cite{stephanov99:_event_by_event_fluct_in,asakawa00:_fluct_probes_of_quark_decon,jeon00:_charg_partic_ratio_fluct_as,ejiri,sasaki,karsch1}.
In particular, higher order cumulants of the event-by-event charge
multiplicity distribution are interesting quantities as  they become
increasingly sensitive to critical behavior near the
phase boundary of QCD
\cite{ejiri,kitazawa,stephanov09:_non_gauss_fluct_near_qcd_critic_point,skokov,stephanov11:_sign_of_kurtos_near_qcd_critic_point,karsch11:_probin_freez_out_condit_in,
Bazavov1,WB_fluc,friman11:_fluct_as_probe_of_qcd}.
In the Beam Energy
Scan (BES) program at the Relativistic Heavy Ion Collider (RHIC) up to
fourth order
cumulants of net-proton fluctuations \cite{STAR_pn_2013}, as a proxy for
net-baryon number fluctuations,
and net-electric charge fluctuations have been measured
\cite{adamczyk14:_beam_energ_depen_of_momen,PHENIX_netcharge}.

While cumulants of conserved charges exhibit singular behavior  at the critical point
\cite{sasaki,stephanov11:_sign_of_kurtos_near_qcd_critic_point}, they
are also  sensitive probes for pseudo-critical behavior in the vicinity
of the chiral crossover transition
\cite{ejiri,karsch11:_probin_freez_out_condit_in,skokov,friman11:_fluct_as_probe_of_qcd,
pn_qm}.
Furthermore,  fluctuations of conserved charges carry  information on
freeze-out conditions  in heavy ion collisions
\cite{karsch11:_probin_freez_out_condit_in,FKfreeze,lgt,lgt1,new, Alba, Kapusta} and they can be
influenced  by different  final state effects
\cite{kitazawa12:_reveal_baryon_number_fluct_from,nahrgang14:_impac_of_reson_regen_and}.

The proximity of the chemical freeze-out, determined by a fit of
statistical hadronization models to data for particle yields in heavy ion
collisions \cite{anton}, and the crossover transition determined in lattice QCD,
suggests that one may gain access to critical phenomena at the QCD phase boundary
by measuring fluctuations of conserved charges at the LHC and in the BES at RHIC
\cite{karsch11:_probin_freez_out_condit_in,lgt}.

In order to identify critical properties of different fluctuations, one needs to
understand a non-critical  baseline for their probability distribution
and resulting cumulants. Note that the often
used ``Poisson baseline'', i.e. the Skellam distribution, is not an
adequate reference for the analysis of net-electric charge and
net-strangeness fluctuations, due to contributions from multi-charged states
and quantum statistics effects.
This is not the case for net-baryon number fluctuations where due
to the absence of multi-charged baryons and the large nucleon mass
a Skellam distribution is indeed an appropriate reference
\cite{braun-munzinger12:_net_charg_probab_distr_in}.

Net-electric charge fluctuations are dominated\footnote{
This is strictly speaking correct for quadratic and quartic charge
fluctuations. We will argue later that baryons and kaons also give
significant contributions to odd order cumulants, e.g. mean and skewness.}
by pions. They are therefore
strongly influenced by quantum statistics effects. Furthermore,
experimental acceptance cuts can severely influence electric charge
fluctuations. Since the Skellam distribution describes the distribution
for the difference of two independent quantities obeying Poisson statistics,
limiting the momentum acceptance does not change the
nature of the distribution. Thus, fluctuations of net baryon number
should be rather weakly influenced by kinematic cuts
\cite{karsch11:_probin_freez_out_condit_in}. However, due
to contributions from multi-charged states and quantum statistic effects,
this is not the case for electric charge fluctuations.

In this study, we discuss the effects of different kinematic cuts
on the net-electric charge fluctuations in a static and expanding
medium. We consider the effect of a transverse momentum cut-off and
of a limited pseudo-rapidity acceptance on different ratios of the $n-$th order
cummulants $\chi_n^Q$  of net-electric charge fluctuations.  Similar
studies,  have been carried out in Ref.~\cite{garg13:_conser}  within  a
hadron resonance gas model.

In the following,  we will focus on the theoretical interpretation of
these results, based on a very transparent example of an ideal gas of
charged pions with quantum statistics, which can be viewed as a
multi-component gas of particles with different charges, masses and
degeneracies.  Such a system contains all relevant properties of
charge fluctuations that allow for a transparent implementation
and interpretation of different kinematic cuts and their consequences.

In particular, we show that the substantial modification of the  higher order cumulant
ratios with momentum cuts , in the acceptance windows of
STAR Collaboration, that have been discussed in Ref.~\cite{garg13:_conser} may be viewed
as arising from the suppression
of multi-charged states, reflecting the Bose momentum distribution in a
pion gas. We will extend this discussion to different low and high
transverse momentum cuts, as well as
pseudo-rapidity windows. We will also focus on properties of the
$\chi_1^Q/\chi_2^Q$ ratio which  is an important baseline to identify
cut-off effects on observable which is  non-critical with respect to
the $O(4)$ universality class.

We perform  a quantitative comparison of our analysis with data obtained by the
PHENIX and STAR Collaborations. However,  in this case we also include
the contribution of other hadronic states using a hadron resonance gas (HRG)
model.  We will show,  that the increase of the lower momentum cut from
$p_{t_\text{min}}=0.2$ GeV (STAR)  to $p_{t_\text{min}}=0.3$~GeV
(PHENIX) does account for a large fraction of the differences present in
the data for mean over variance, $M_Q/\sigma_Q^2$. Similar differences
are expected to show up in data for the  normalized kurtosis,
$\kappa_Q\sigma_Q^2$.


Finally we will discuss the relation between non-zero momentum
cut-off and finite volume effects in a non-interacting pion gas.
This allows to argue that lattice QCD calculations of electric charge
fluctuations performed in a finite volume are actually adequate
to describe charge fluctuations determined experimentally and
infinite volume extrapolation may even not be needed due to the
similarity between finite volume and non-zero transverse momentum
cut-off effects.

The paper is organized as follows: In the next section
we discuss the  momentum  cut-off dependence of electric charge fluctuations
in a pion gas.  In Section III we introduce STAR and PHENIX data on cumulants
of charge fluctuations $\chi_n^Q$  and their interpretation. In Section IV the
effects of an expanding medium on fluctuations will be discussed.
In Section V we compare the effects of finite size and momentum cuts on
$\chi_n^Q$. Our conclusions are presented in Section VI.

\section{Charge fluctuations in a pion gas}

\label{sec:pifluc}

The thermodynamic pressure for a non-interacting pion gas at
temperature $T$ and electric charge chemical potential $\mu_Q$ is given by
\begin{equation}
 p(T,\mu_Q) = -T\sum_{i=1}^{3} \int \frac{d^3 \boldsymbol{p}}{(2\pi)^3}
  \ln{(1-e^{-(E_p-\mu_Q Q_i)/T})}\label{eq:pressure}
\end{equation}
with
\begin{equation}
 \int d^3 \boldsymbol{p} = 2\pi
 \intop_{\eta_{\text{min}}}^{\eta_{\text{max}}} d\eta
 \intop_{p_{t_{\text{min}}}}^{p_{t_\text{max}}}dp_t p_t |\boldsymbol{p}|\label{eq:detadpt}
\end{equation}
where  $E_p=\sqrt{|\boldsymbol{p}|^2+m^2}$ is the pion energy  and  $Q_i=0,\pm 1$
denotes their electric charge.

The momentum integration in Eq.~\eqref{eq:detadpt}, is expressed
 in terms of the transverse momentum $p_t$ and pseudo-rapidity
$\eta =  [\ln{(|\boldsymbol{p}|+p_z)/(|\boldsymbol{p}|-p_z)}]/2$.
We consider the  experimental acceptance
window, $0.2~{\rm GeV} \le p_t \le 2$~GeV and $|\eta| < 0.5$  for STAR
\cite{adamczyk14:_beam_energ_depen_of_momen},  and
$0.3~{\rm GeV} \le  p_t \le 2$~GeV and
$|\eta| < 0.35$ for PHENIX measurements \cite{PHENIX_netcharge}.

The cumulants of electric charge fluctuations are obtained from
\begin{equation}
 \chi_n^Q \equiv \frac{\partial^n}{\partial(\mu_Q/T)^n}
\frac{p(T,\mu_Q)}{T^4} \; .
\label{eq:chi_n-pressure}
\end{equation}
To characterize the
properties  of the system, it is convenient to introduce different
ratios of $\chi_n^Q/\chi_m^Q$. Hereafter we focus on two ratios, the mean
over variance, $M_Q/\sigma_Q^2 = \chi_1^Q/\chi_2^Q$ and the normalized
kurtosis $\kappa_Q\sigma_Q^2 = \chi_4^Q/\chi_2^Q$.

In the presence of a non-zero momentum cut-off the cumulants
may be represented by the standard cluster expansion
\begin{eqnarray}
  \chi_n^Q &=& \frac{m^2}{\pi^2 T^2}\sum_{k=1}^{\infty}k^{n-2}\hat K_2(k m/T )\cosh{( k\mu_Q/T)},\ n={\rm even}
\nonumber \\
  \chi_n^Q &=& \frac{m^2}{\pi^2 T^2}\sum_{k=1}^{\infty}k^{n-2}\hat K_2(k m/T )\sinh{( k\mu_Q/T)},\ n={\rm odd}
\nonumber \\
 \label{expansion}
\end{eqnarray}
where the function $\hat K_2(x)$ is defined as,
\begin{align}
 \label{exp1}
 \hat K_2(km/T) &= \frac{k}{2m^2 T}\intop_{\eta_{\text{min}}}^{\eta_{\text{max}}}
 \!\!d\eta \!\!
 \intop_{p_{t_{\text{min}}}}^{p_{t_\text{max}}}\!\!dp_t p_t
 |\boldsymbol{p}| e^{-k\sqrt{|\boldsymbol{p}|^2+m^2}/T}.
\end{align}

It reduces to the modified Bessel function $K_2(km/T)$ after
integration  over  the  entire momentum space,
$-\infty < \eta < \infty$ and $0 < p_t < \infty$.
This makes the influence of the Bose statistics very transparent and highlights
the similarity between cut-off effects on Bose particles and multi-charged hadrons.

\begin{figure}[t]
 \centering
 \includegraphics[width=3.375in]{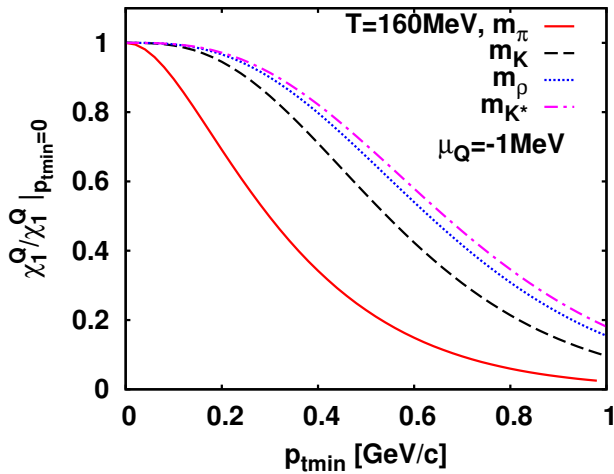}
 \caption{The lower transverse momentum cut-off dependence of the first
 order cumulant $\chi_1^Q$ of the electric charge fluctuations in a Bose gas at different
 values of the particle mass, $m=m_\pi, m_K, m_\rho$ and $m_{K^*}$ at $T=160$~MeV and
 $\mu_Q=-1$~MeV. The results are normalized to the corresponding values at $p_{t_{\text{min}}}=0$.}
 \label{fig:chi1Q}
\end{figure}

\begin{figure}[t]
 \centering
 \includegraphics[width=3.375in]{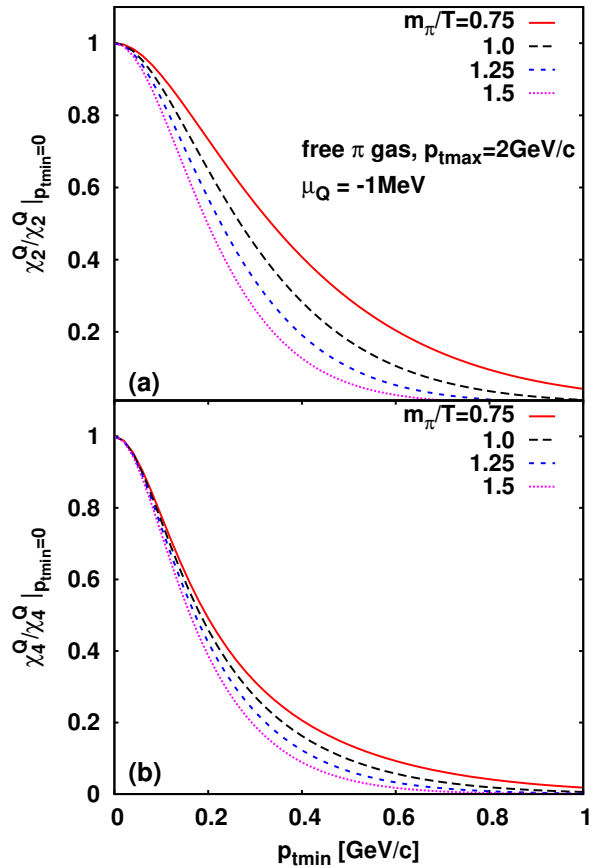}
 \caption{The lower transverse momentum cut-off dependence of the quadratic
  [$\chi_2^Q$, (a)] and quartic [$\chi_4^Q$, (b)] cumulants of the electric charge fluctuations
 in a pion gas at different values of the temperature, $100~{\rm MeV}\ \lsim\ T\ \lsim\ 200~{\rm MeV}$
and for $\mu_Q=-1$~MeV. The results are normalized to the corresponding
 values obtained  at $p_{t_{\text{min}}}=0$.}
 \label{fig:cutoffdependence}
\end{figure}
From Eq.~\eqref{expansion} it is clear that in a Bose gas
$\chi_n^Q$  can be regarded as a sum over 
contributions from multi-charged
particles with mass $km$, charge $k$ and degeneracy $k^{n-4}$,  where
$k=1,2,\cdots, \infty$. The  $k=1$ term corresponds to the Boltzmann
approximation for which cumulant ratios $\chi_{n+2}^Q/\chi_n^Q$=1,
i.e., the normalized kurtosis $\kappa_Q\sigma^2_Q=1$. For $k=1$ 
the corresponding probability distribution  $P(N)$ for net charge $N$  
is the Skellam
function \cite{BraunMunzinger:2011dn,Skellam}. Contributions of
multi-charged Boltzmann particles with $k>1$,
result in a wider $P(N)$,
which has a broader tail than the Skellam function
\cite{braun-munzinger12:_net_charg_probab_distr_in}.
This leads to larger quartic ($\chi_4^Q$) than quadratic ($\chi_2^Q$)
charge fluctuations. More generally, the higher order terms in
$k$, which become increasingly important for large $n$ owing to the degeneracy
factor $k^{n-4}$,  imply that $\chi_{n+2}^Q > \chi_{n}^Q$.
Consequently, in the presence of multi-charged particles, as well as particles
with Bose statistics,  the normalized kurtosis $\kappa_Q\sigma^2_Q$  is always
larger than unity in a gas of non-interacting hadrons.
In Eq. (\ref{expansion}) the momentum cut-off dependence enters 
through the function $\hat{K}_2(km/T)$ where the 
effect of $p_{t_{\text{min}}}\neq 0$ quantitatively depends on $m$, $T$ and $k$ 
separately. 
A non-zero $p_{t_{\text{min}}}$ thus 
modifies the relative contribution of different 
$k$-terms  to $\chi_n^Q$.

Fig.~\ref{fig:chi1Q} displays the $p_{t_\text{min}}$ dependence
of the first cumulants at $T=160$~MeV and $\mu_Q=-1$~MeV for various
particle masses, $m=m_\pi, m_K, m_\rho$ and $m_{K^*}$, normalized to their 
values calculated with
$p_{t_\text{min}}=0$.

The charge chemical potential is assumed  to be
negative and small, as expected in heavy ion collisions at high energies.
For $m=m_{K^*}$, the mass is already so large that the quantum
statistics effect is almost negligible,  thus the result can be regarded as that
of the Boltzmann approximation where the higher $k$-terms in Eq. \eqref{expansion}
are neglected. 

By decreasing the particle mass, one sees a stronger
dependence of $\chi_1^Q$ on $p_{t_{\text{min}}}$, reflecting the
contribution from $k>1$ terms which are also understood as
the enhancement of the particle number at low $p_t$ in the Bose
distribution. 

In Fig.~\ref{fig:cutoffdependence} we show the cut-off dependence
of the two even order cumulants $\chi_2^Q$ and
$\chi_4^Q$ for a pion gas at $\mu_Q=-1$~MeV and at different
temperatures.
It is clear that cut-off effects are more severe for higher order
moments and they are more pronounced at lower temperature.
This temperature dependence
can be 
deduced from
the $\hat{K}_2(km/T)$ function by changing the integration variable
from $p_t$ to $x=kp_t/T$,
which shifts the lower limit of the integration range to
$x_{\text{min}} = kp_{t_\text{min}}/T$.
An increase of the temperature,  therefore effectively lowers the $p_t$
cut. Consequently, the cumulants at higher temperature take a value
closer to those without $p_t$ cut. 

\begin{figure}[t]
 \centering
 \includegraphics[width=3.375in]{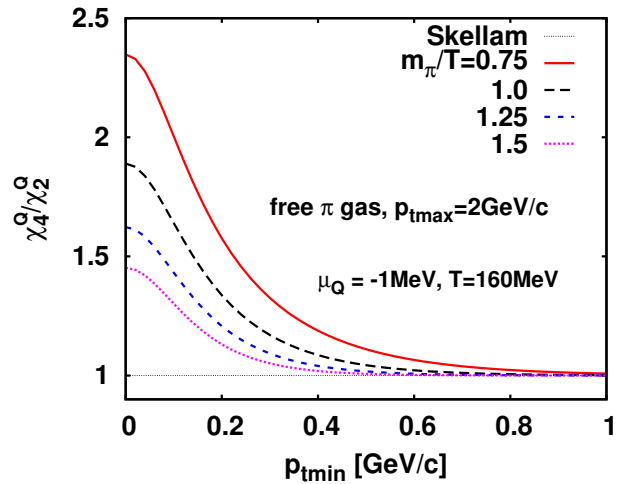}
 \caption{The lower transverse   momentum cut-off dependence of the ratio of quartic and quadratic electric
charge fluctuations in a pion gas at different values of the temperature,
$100~{\rm MeV}\ \lsim\ T\ \lsim\ 200~{\rm MeV}$
for $\mu_Q=-1$~MeV.
}
 \label{fig:c4c2_cutoffdependence}
\end{figure}

The difference in the cut-off
dependence of $\chi_2^Q$ and $\chi_4^Q$  will  also influence their ratio,
$\chi_4^Q/\chi_2^Q=\kappa_Q\sigma_Q^2$. However, the effect is weaker in
$\kappa_Q\sigma_Q^2$,  due to a  partial cancelation  of a similar dependence
of quartic and quadratic fluctuations on $p_t$ momentum cut-off.
This is evident in  Fig.~\ref{fig:c4c2_cutoffdependence}, where we show the momentum cut-off dependence of $\chi_4^Q/\chi_2^Q$. 
The temperature dependence
of $\kappa_Q\sigma_Q^2$ indicates, that there
are stronger cut-off effects at higher temperature, where the
multi-charged states in the cluster sum become increasingly important.

The above discussion of the   
temperature
dependence of the  cumulants and their ratios
at different $p_t$ cuts can be  extended to more general cases. For
$|\mu_Q| \ll T$ and $|\mu_Q| \ll m$,   which is  the case in heavy ion collisions at
RHIC and LHC energies,  the cluster expansion  Eq.~\eqref{expansion} implies, that 
\begin{equation}
 \chi_{2n+1}^Q/\chi_{2n}^Q \simeq \mu_Q/T\label{eq:chi1chi2_pi} \;\; ,
\end{equation}
irrespective of the $p_t$ cut. This is because, the $\hat{K}_2(km/T)$ functions
cancel in these ratios. Consequently, in the pion gas,  the lowest ratio $\chi_1^Q/\chi_2^Q$ does not
depend on the $p_t$-cut. Similarly, one finds
\begin{equation}
 \frac{\chi_{2n+1}^Q}{\chi_{2n-1}^Q} \simeq \frac{\chi_{2n+2}^Q}{\chi_{2n}^Q},
\end{equation}
which implies that  $\chi_4^Q/\chi_2^Q \simeq \chi_3^Q/\chi_1^Q$.
These relations hold
exactly for $\mu_Q=0$, but deviations at $\mu_Q \neq 0$, in the range of RHIC and LHC energies,  are
found to be less than 1\%. 

\begin{figure}[t]
 \centering
 \includegraphics[width=3.375in]{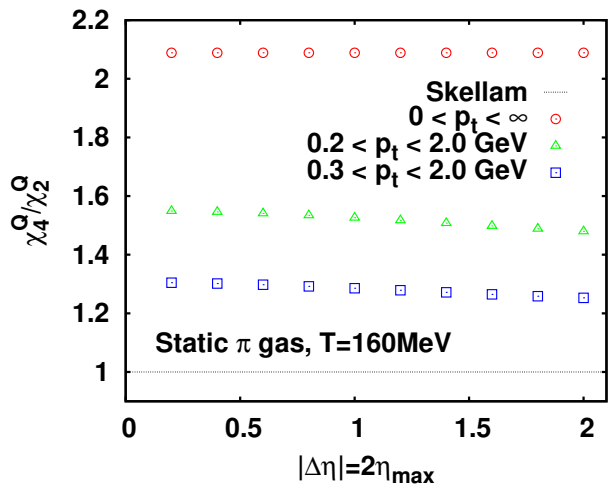}
 \caption{Normalized kurtosis $\kappa_Q\sigma_Q^2=\chi_4^Q/\chi_2^Q$ for an ideal
pion gas at $T=160$~MeV, $\mu_Q=0$ and in  different transverse momentum windows. Open circles
stand for the fully integrated $p_t$ range. Open triangles  and
squares show the result with the same $p_t$ acceptance as STAR and PHENIX
 measurements, respectively \cite{adamczyk14:_beam_energ_depen_of_momen,PHENIX_netcharge}.}
 \label{fig:statickurtosis}
\end{figure}

Fig.~\ref{fig:statickurtosis} shows the  normalized kurtosis,
$\kappa_Q\sigma_Q^2$,   calculated in a pion gas at $T=160$ MeV
as a function of the pseudo-rapidity cut $\eta_{\text{max}}$, and
for different  $p_t$ cuts corresponding to the STAR and PHENIX
experiments, respectively.

From Figs.~\ref{fig:cutoffdependence}-\ref{fig:statickurtosis} we
conclude that the
dependence of electric charge cumulants on transverse momentum cuts is strong while the
dependence on different pseudo-rapidity cuts is weak. We also checked the dependence of
cumulant ratios on the high momentum cut-off, $p_{t_\text{max}}$,  and found that this is
negligible for $p_{t_\text{max}}> \ 1$~GeV. (see Sec.~\ref{sec:expansion})

The difference  in the dependence of $\kappa_Q\sigma_Q^2$ on the transverse momentum
and pseudo-rapidity cuts is transparent from the decomposition of the
Bose integral in terms of the multi-charged cluster sum. A transverse momentum
cut-off,  $p_{t_\text{min}}$,  increases the effective mass of  particles,
$m_{\text{eff}}=\sqrt{m^2+ p^2_{t_\text{min}}}$, and thus further suppresses the statistical
weight of the $k$-cluster contribution relative to the leading Boltzmann term.
Consequently, increasing $p_{t_\text{min}}$ reduces
$\kappa_Q\sigma_Q^2$ towards unity, which is its value for the Boltzmann
statistics.
Cuts in pseudo-rapidity, on the other hand, do not restrict the minimal
momentum in the particle dispersion relation and therefore do not lead
to an additional suppression of multi-charged clusters contributing to
$\kappa_Q\sigma_Q^2$.

\section{ Electric charge fluctuations in the PHENIX and STAR data}

We have shown in the previous section, that a cut-off imposed on the low momentum part of the particle phase-space
can have significant effects on different  cumulants of net electric charge fluctuations and their ratios.
Thus, the difference of $100$~MeV
in the transverse momentum cut-off used by the PHENIX and STAR Collaborations  \cite{adamczyk14:_beam_energ_depen_of_momen,PHENIX_netcharge}, should naturally
imply differences in their results on cumulants of electric charge fluctuations.

However, in order to quantify the influence of momentum cut-off effects
on charge fluctuations in the context of heavy ion data, one needs to  extend
the pion gas calculation by including contributions from all known
charged hadrons and resonances. In the following, we will apply a hadron
resonance  gas (HRG) model which is very successful
in the description of the equation of state and different fluctuation
observables calculated in lattice QCD in the hadronic phase
\cite{Bazavov2,karsch,Ratti}.

We first consider the
influence of a non-zero $p_{t_\text{min}}$ on the ratio of mean and variance
of net-electric charge fluctuations, $M_Q/\sigma_Q^2=\chi_1^Q/\chi_2^Q$. This
quantity is statistically well under control and a direct
comparison between data obtained by STAR and PHENIX, as well as with model
calculations, is meaningful.

The ratio $M_Q/\sigma_Q^2$ involves the first order
cumulant $\chi_1^Q$, which receives significant contributions from
strange mesons as well as charged baryons.
In fact,
in the case relevant for heavy ion collisions, i.e. in a strangeness neutral system,
$M_S=0$, with a fixed ratio of net-electric charge and net-baryon number, $M_Q/M_B\simeq 0.4$,
the electric charge chemical potential required to fulfill these constraints is
negative. The contribution of charged, non-strange mesons to $M_Q$ thus is negative
and the mean becomes positive only due to contributions from charged baryons and strange
mesons. Although in a single component Bose gas
the ratio $M_Q/\sigma_Q^2$ does not depend on
the $p_t$ cut-off, Eq.~\eqref{eq:chi1chi2_pi},
the cumulants themselves are modified depending on the mass of this
component, as seen in Fig.~\ref{fig:chi1Q}. However, in a multi-component system, 
the change of $M_Q$ and
$\sigma_Q^2$ of each hadronic contribution does not cancel anymore in their ratio.

We have calculated $M_Q/\sigma_Q^2$ in a HRG model using three different scenarios for
introducing the low momentum cut-off: (A)  $p_{t_\text{min}}$
 is introduced only  for the pion contribution, (B) an identical  cut-off is introduce  for all charged hadrons,  and (C) a larger cut-off, $p_{t_\text{min}}=0.4$~GeV, is introduced
for protons and anti-protons,  as it is done in the electric charge measurements of
the STAR Collaboration.

\begin{figure}[t]
 \centering
 \includegraphics[width=3.375in]{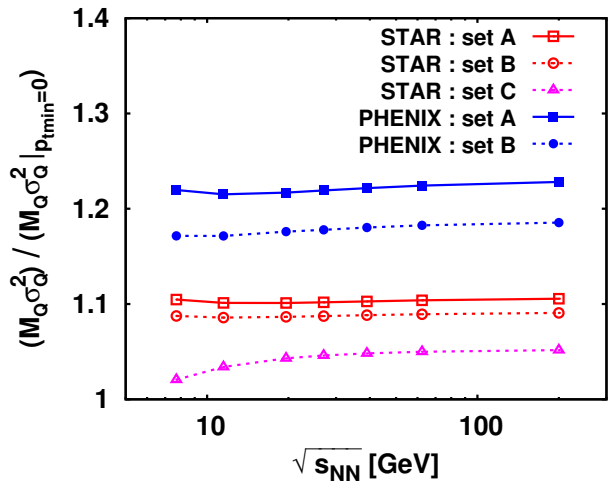}
 \caption{The ratio  of mean and variance, $M_Q/\sigma^2_Q$, of electric
 charge fluctuations  for the three different scenarios (\emph{see text})
 with STAR
 acceptance (open symbols) and PHENIX acceptance (closed
 symbols) in a HRG model. The energy dependence of thermal parameters at chemical freeze-out
 are taken from Ref.~\cite{karsch11:_probin_freez_out_condit_in}. The
 results are normalized with the values for $p_{t_{\text{min}}}=0$.}
 \label{fig:R12}
\end{figure}

We show results from this analysis in Fig.~\ref{fig:R12},
where $M_Q/\sigma_Q^2$ is calculated along the freeze-out line  in central heavy ion collisions,
with the energy dependent thermal parameters, $\mu_B$, $\mu_Q$,  $\mu_S$ and $T$
 from  Ref.~\cite{karsch11:_probin_freez_out_condit_in}. 
We find that the effect of a low momentum cut-off on
$M_Q/\sigma_Q^2$ is almost independent of the collision energy.
Comparing the different low momentum cut-off scenarios, one finds that
the set (A) gives the largest effect, since it corresponds to
cutting off the negative contribution of pions to $M_Q$ and the
pion contribution to $\sigma_Q^2$ is reduced. Incorporating the cut-off
for other hadrons reduces the changes from the values without the
cut-off. This is because the kaon contribution partly compensates the
increase of $M_Q$ due to the change of the pion contribution.
The result of set (C) shows a weak $\sqrt{s_{NN}}$ dependence below
$\sqrt{s_{NN}} \leq 20$~GeV, due to the dominance of the proton contribution.
The difference among  different
scenarios is smaller than that observed between STAR and PHENIX acceptance.

We conclude that a non-zero transverse momentum cut-off
increases the ratio $M_Q/\sigma_Q^2$ over the value obtained without any
cuts. The larger $p_{t_\text{min}}$ used by PHENIX suggests that their  $M_Q/\sigma_Q^2$
ratio  is about 10\% larger than the
STAR values. This accounts for about $1/3$ of the difference between the
PHENIX and STAR data. The ratio of these two data sets is, within large
errors, constant over the entire energy range. A fit yields
$(M_Q/\sigma_Q^2)_\text{PHENIX}/(M_Q/\sigma_Q^2)_\text{STAR} =
1.35(5)$. Note, however, that more than 99\% of the error quoted
by PHENIX is systematic.

\begin{figure}[!t]
 \centering
 \includegraphics[width=3.375in]{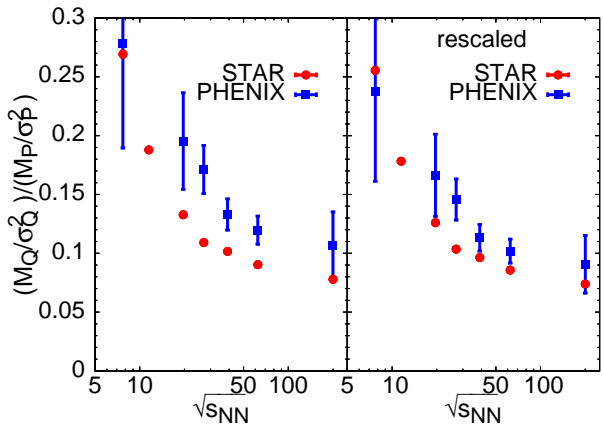}
 \caption{The ratio  of mean and variance, $M_Q/\sigma^2_Q$, of electric charge fluctuations  obtained by STAR (circle)
and PHENIX (squares) Collaboration  at different beam energies \cite{adamczyk14:_beam_energ_depen_of_momen,PHENIX_netcharge},  (left panel).
Data have been normalized to the corresponding ratio for net-proton
number fluctuations, $M_P/\sigma_P^2$, obtained by STAR \cite{STAR_pn_2013}. The right
hand panel shows the rescaled data
$(M_Q/\sigma_Q^2)_{p_{t_\text{min}}=0}= r (M_Q/\sigma_Q^2)_{p_{t_\text{min}}}$.
Where the rescaling factor $r$ is taken from the result of set (C) for the STAR data and set (B) for the PHENIX data shown
in Fig.~\ref{fig:R12}.
}
 \label{fig:MQsQ2}
\end{figure}

In Fig.~\ref{fig:MQsQ2} we show the data measured by STAR  \cite{adamczyk14:_beam_energ_depen_of_momen} and PHENIX \cite{PHENIX_netcharge}.
They have been normalized by the corresponding ratio for net-proton
fluctuations measured by STAR
\cite{STAR_pn_2013}. As can be seen from the right hand panel of
Fig.~\ref{fig:MQsQ2},  rescaling the STAR and PHENIX data to vanishing transverse
momentum cut-off values removes a large part of the differences between these
data. In fact, within the large systematic errors of the PHENIX data it
yields consistent results.

\begin{figure}[t]
 \centering
 \includegraphics[width=3.375in]{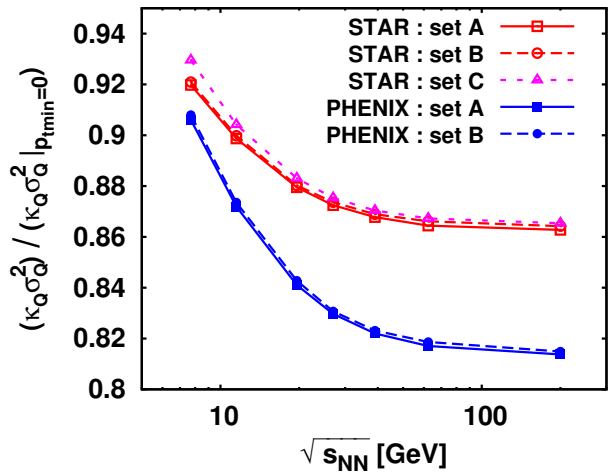}
 \caption{Normalized kurtosis $\kappa_Q \sigma^2_Q$ for the three
 difference cut-off scenarios in a HRG model along the freeze-out line. The legends are
 the same as in Fig.~\ref{fig:R12}.}
 \label{fig:kappasigma2_sqrts}
\end{figure}

In Fig.~\ref{fig:kappasigma2_sqrts} we show the influence of a non-zero
transverse momentum cut-off on the normalized kurtosis,
$\kappa_Q\sigma_Q^2$. Compared to the pion gas results shown in
Fig.~\ref{fig:c4c2_cutoffdependence},
the reduction of $\kappa_Q \sigma^2_Q$ is
smaller and turns out to be 10-20\%
due to a smaller cut-off dependence of the heavier additional degrees of
freedom in a HRG. The small discrepancy in the different scenarios
nevertheless implies that the main reason for
the reduction is the pion
effect. Contrary to $M_Q/\sigma_Q^2$, this result
shows a stronger dependence of $\kappa_Q \sigma_Q^2$ on
$\sqrt{s_{NN}}$. Thus, we expect
that the difference in $p_{t_{\text{min}}}$ between STAR and PHENIX
acceptance will lead to about 6\% difference in $\kappa_Q \sigma_Q^2$ at $\sqrt{s_{NN}}=200$ GeV,
and it will be negligible at $\sqrt{s_{NN}}=7.7$ GeV. We note, that
similar results have been obtained in Ref.~\cite{garg13:_conser} where a
further reduction of $\kappa_Q\sigma_Q^2$ due to effects of resonance
decays is also discussed.

\section{Effects of expansion}
\label{sec:expansion}

\begin{figure}[!t]
 \centering
 \includegraphics[width=3.375in]{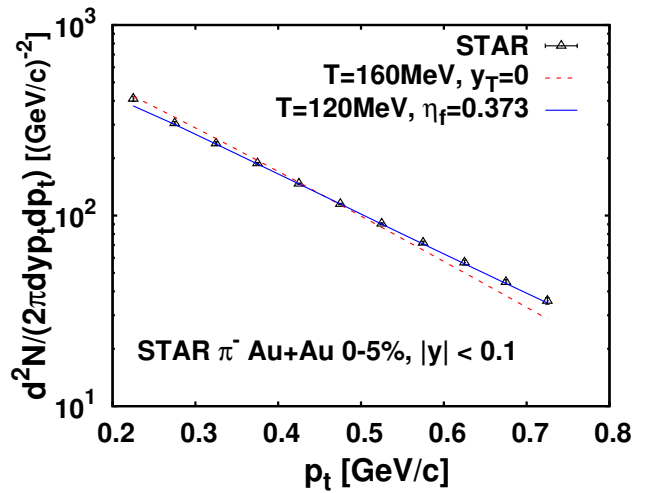}
 \caption{The $p_t$ distribution of charged pions. The spectrum of the
 STAR data \cite{abelev09:_system_measur_of_ident_partic} is shown as
 triangles.  The dashed-line is the model results at chemical freeze-out
 ($T=160$~MeV) from a longitudinally expanding fireball. The
 solid-line  represents the distribution  at thermal freeze-out for
 a longitudinally and transversally  expanding
 fireball with parameters fitted to the STAR data \cite{abelev09:_system_measur_of_ident_partic}. }
 \label{fig:pt}
\end{figure}

Until now, we have discussed the influence of momentum cuts imposed in a
thermal medium at rest. In  heavy ion experiments, however, the kinematic
cuts on measured fluctuations are applied to hadrons that move
freely only after kinetic
freeze-out. Consequently, for the comparison of experimental data
with theoretical calculations, it is important
to verify how momentum
cuts implemented in different stages of the fireball evolution can influence
the measurement of fluctuations of conserved charges \cite{garg13:_conser}.

Here we employ an expanding source model with parameters chosen such that
the pion $p_t$ spectrum measured by the STAR Collaboration
\cite{abelev09:_system_measur_of_ident_partic} is reproduced. We compare the effect
of a $p_t$ cut imposed either at the time of chemical or thermal freeze-out on the
measured values for the kurtosis of net-electric charge fluctuations.

Assuming boost invariance along the collision axis and transverse
expansion with a linear rapidity profile, as well as an instantaneous emission at
$\tau=\tau_f$ hyper-surface,
we parametrize the  phase space distribution of  pions as
\cite{chapman95:_thermalsourcemodel}
\begin{equation}
 S(x,\boldsymbol{p}) = \frac{m_T
  \cosh(y-\eta_s)\delta(\tau-\tau_f)}{(2\pi)^3}n_B(u\cdot p)e^{-\frac{x^2+y^2}{2R^2}}.\label{eq:sourcefunc}
\end{equation}
Here $y = \ln[(E_p+p_z)/(E_p-p_z)]/2$ is the rapidity of the pions,
$\eta_s= \ln[(t+z)/(t-z)]/2$ is the space-time rapidity which is equal to the
longitudinal flow rapidity $y_L$,
and $R$ is the Gaussian transverse size of an expanding fireball. The
flow 4-velocity is parametrized as
\begin{align}
u^\mu = (\cosh y_T\cosh \eta_s ,\sinh
y_T\cos\phi, \sinh y_T\sin\phi,\\ \cosh y_T \sinh \eta_s).\nonumber
\end{align}
The transverse flow rapidity is assumed to be $y_T = \eta_f r/R$,
with $r=\sqrt{x^2+y^2}$ and a
flow control parameter $\eta_f$ which is adjusted so as to reproduce
the $p_t$ spectrum of pions.

 Since the invariant spectrum is obtained from the
source function from  Eq.~\eqref{eq:sourcefunc}, as
\begin{equation}
 E_p\frac{dN}{d^3 \boldsymbol{p}} = \int d^4x S(x,\boldsymbol{p}),
\end{equation}
the $p_t$ spectrum can be obtained by integrating over space-time
variables,  as
$\int d^4x = 2\pi \tau_f \int d\tau \int d\eta_s \int r dr$.
Consequently,  cumulants of charge  fluctuations are given by
\begin{equation}
 \label{chie}
  \chi_n^Q \propto \frac{\partial^{n-1}}{\partial \mu_Q^{n-1}} \int
  \frac{d^3\boldsymbol{p}}{E_p}S(x,\boldsymbol{p}).
\end{equation}
The proportionality factor is canceled out when taking ratios such as
$\chi_4^Q/\chi_2^Q$. Note that Eq.~\eqref{chie} reduces to
Eq.~\eqref{eq:chi_n-pressure} when considering a static source, i.e.
by choosing $y_L=0$ and $y_T=0$.

\begin{figure}[!t]
 \centering
 \includegraphics[width=3.375in]{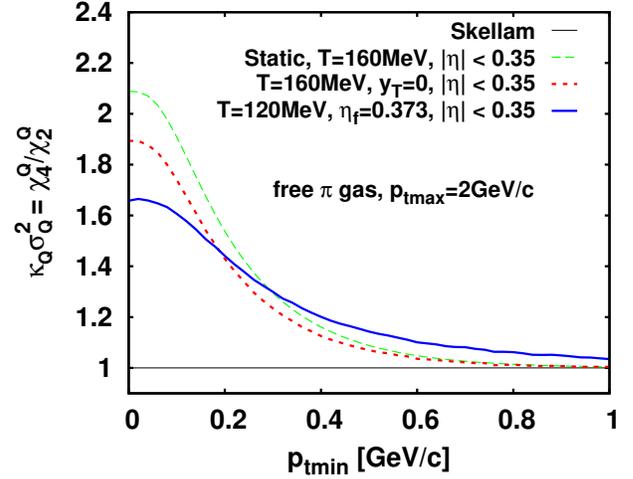}
 \caption{The kurtosis $\kappa_Q \sigma_Q^2=\chi_4^Q/\chi_2^Q$ as a
 function of $p_{t_{\text{min}}}$. The long-dashed line
 stands for the static source from Eq. ~\ref{eq:pressure}. The short-dashed and
 solid lines denote the results obtained from the spectra shown in Fig.~\ref{fig:pt}. }
 \label{fig:ptcut}
\end{figure}

Fig.~\ref{fig:pt} displays the $p_t$ spectrum of $\pi^-$ measured by
the STAR Collaboration \cite{abelev09:_system_measur_of_ident_partic}. We fit the data with the above source model with
$T=120$~MeV and $R=6$~fm, and obtain $\eta_f = 0.373$. Although the overall scale factor
is also fitted, our result  still underestimates the yields at the lowest
$p_t$ bin. For comparison we also plot the spectrum at  $T=160$~MeV
without transverse flow, i.e. for $y_T=0$ but at finite $y_L$. One sees
that it does not reproduce the slope of the data. Therefore, it is expected
that the influence of a $p_t$ cut on fluctuation observables can be different
at chemical and thermal freeze-out due to differences in the $p_t$
spectrum.

Fig.~\ref{fig:ptcut} shows the normalized kurtosis $\kappa_Q
\sigma^2_Q$ calculated in an
 expanding  and static source with and without transverse flow for
different $p_{t_\text{min}}$. The
pseudo-rapidity  window  is fixed to $|\eta|<0.35$  and the upper
momentum cut  to $p_{t_\text{max}}=2$~GeV.  For
$p_{t_\text{min}}<0.2$~GeV
there is a clear reduction of $\chi_4^Q/\chi_2^Q$ relative to its value obtained
with a static source from Eq.~\eqref{eq:pressure}, when the longitudinal
and/or transverse expansion is included.  However, for $p_{t_\text{min}}\geq 0.2$~GeV
the effect of a $p_t$-cut on $\kappa_Q\sigma^2_Q$ is similar in all cases.
The values of the kurtosis do not differ much for
$p_{t_\text{min}} \geq 0.2$~GeV and the  $p_{t_\text{min}}$
dependence is only slightly modified in the presence of transverse
flow. As can be seen in Fig.~\ref{fig:pt} the $p_t$ spectrum at thermal
freeze-out, $T=120$~MeV and $\eta_f=0.373$,  clearly has a larger slope
than that calculated at $T=160$~MeV without transverse flow. This difference
in the spectral shapes leads to a somewhat slower convergence of
$\kappa_Q\sigma^2_Q$  to unity with increasing  $p_{t_\text{min}}$.
Nonetheless, the resulting decrease of $\kappa_Q \sigma^2_Q$ in the interval
$p_{t_\text{min}} \in [0.2~{\rm GeV},0.3~{\rm GeV}]$ is as large as that
found for the static source.

\begin{figure}[!t]
 \centering
 \includegraphics[width=3.375in]{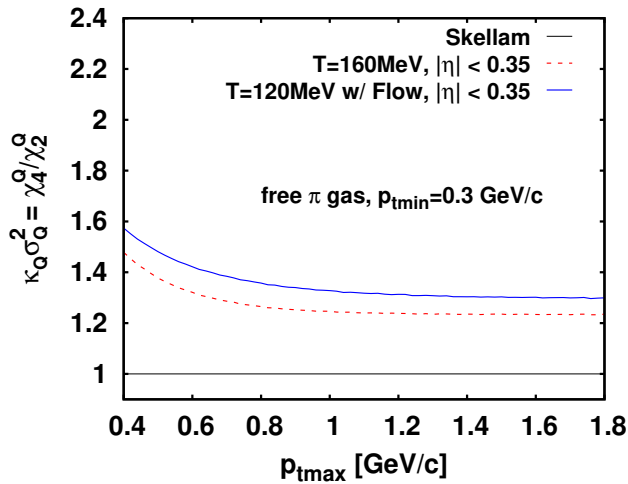}
 \caption{Dependence of the ratio of  quartic ($\chi_4^Q$) and quadratic ($\chi_2^Q$)  cumulants of
 electric charge fluctuations  on
 $p_{t_\text{max}}$ for fixed  $p_{t_\text{min}}=0.3$~GeV and
 $|\eta|=0.35$. The results correspond to pion spectra shown in
 Fig.~\ref{fig:pt}. }
 \label{fig:ptmax}
\end{figure}

For completeness we also show in Fig.~\ref{fig:ptmax}  the influence of
$p_{t_\text{max}}$  on $\kappa_Q\sigma^2_Q$ at fixed  $p_{t_\text{min}}=0.3$~GeV
and $|\eta|<0.35$.  There is no notable change of the kurtosis when
decreasing  $p_{t_\text{max}}$ from 2~GeV  to 1~GeV. For
$p_{t_\text{max}}<\ 1$~GeV,   the relative contribution of lower momentum
particles with charges $k>1$ in Eq.~\eqref{expansion}, i.e. higher order
terms appearing due to Bose statistics,  start to dominate
fluctuations, consequently $\kappa_Q\sigma^2_Q$ is increasing with
decreasing $p_{t_\text{max}}$ . The dependence of $\kappa_Q\sigma^2_Q$ on
$p_{t_\text{max}}$ is similar for a static and transversally expanding
medium.

The above results on flow effects were derived in a rather simplified
system of a non-interacting pion gas. However, due to leading
contributions of pions to electric charge fluctuations and a large mass
of multi-charged baryons, our conclusions  should  also hold  in a
hadron resonance gas which accounts for contributions of all known
hadrons and resonances.

The sensitivity to kinematic cuts may still be
modified if charge fluctuations are influenced  by the critical chiral
dynamics \cite{morita},  which is clearly not included in this study.
For the case of charge fluctuations in the crossover region, however,
we conclude that flow effects can safely be ignored in the analysis
of charge fluctuations.

\section{ Transverse momentum cut-off and finite-size effects}

In the previous sections we have seen that a quantitative comparison
between experimental data taken in specific acceptance windows, in
particular with a non-zero transverse momentum cut-off, with the thermal
resonance gas model requires also to take into account corresponding cut-offs
in the model calculation.

In a non-interacting hadron gas a
non-zero $p_t$ cut reduces the available phase space, which may be viewed
as an increase of the effective mass of particles, and thus leads to
corresponding changes in thermodynamic observables. This effect is
similar to the thermodynamics of particles in a finite volume
\cite{EKS,Bhattacharyya:2015zka}. The finite size of a system, e.g. a
box of size $L^3$, suppresses low momentum states.

 When considering a box with periodic boundary conditions the lowest
non-zero momentum of bosons in such a box is $2\pi/L$. In the case
of a non-interacting scalar field theory, which describes the thermodynamics
of spin zero bosons, i.e. pions, it is straightforward to analyze the
volume dependence of thermodynamic observables \cite{EKS}. In particular, one
can calculate the volume dependence of conserved charge fluctuations and
compare the influence of finite volume effects with those of a transverse
momentum cut-off.

\begin{figure}[!t]
 \centering
 \includegraphics[width=3.375in]{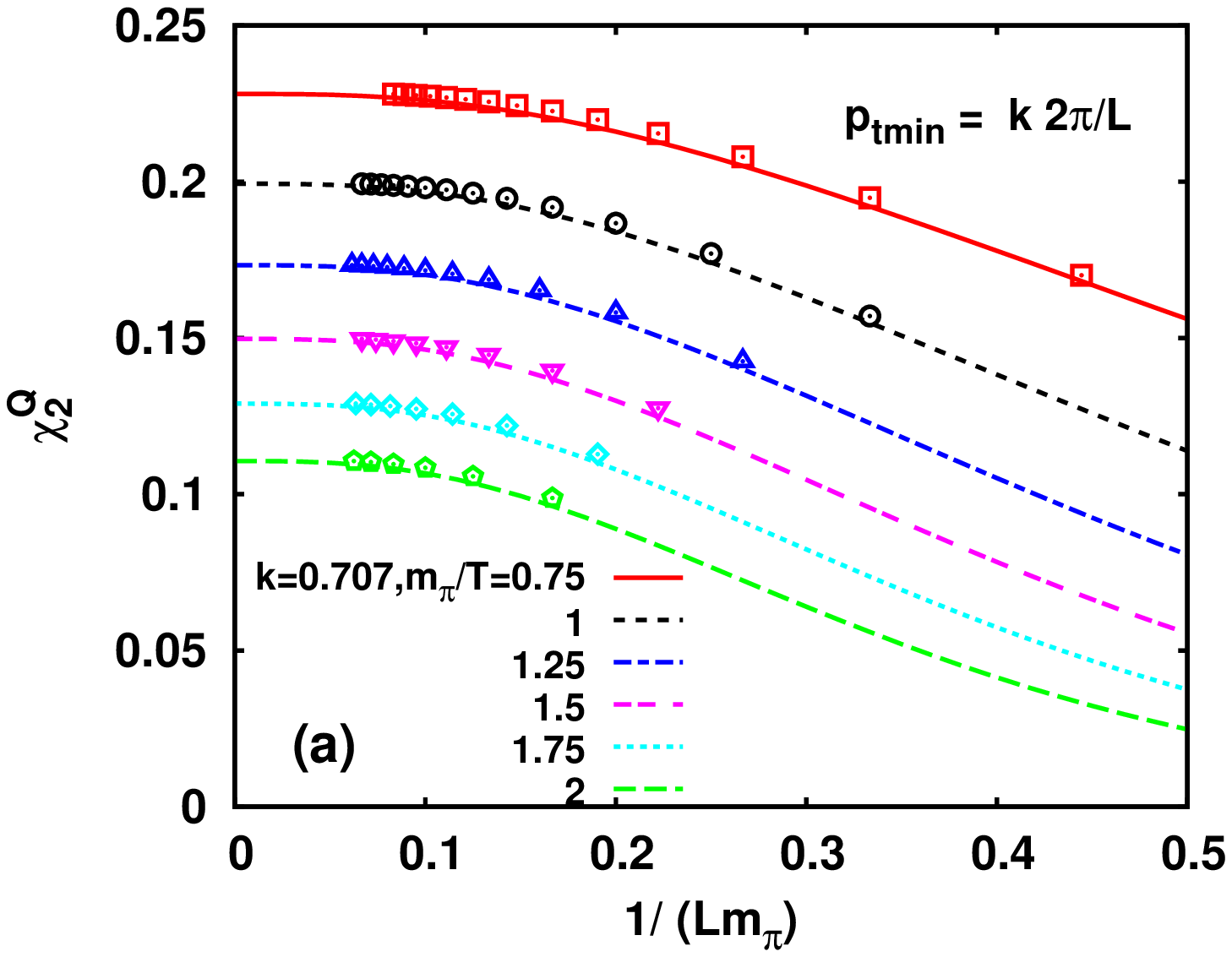}
 \includegraphics[width=3.375in]{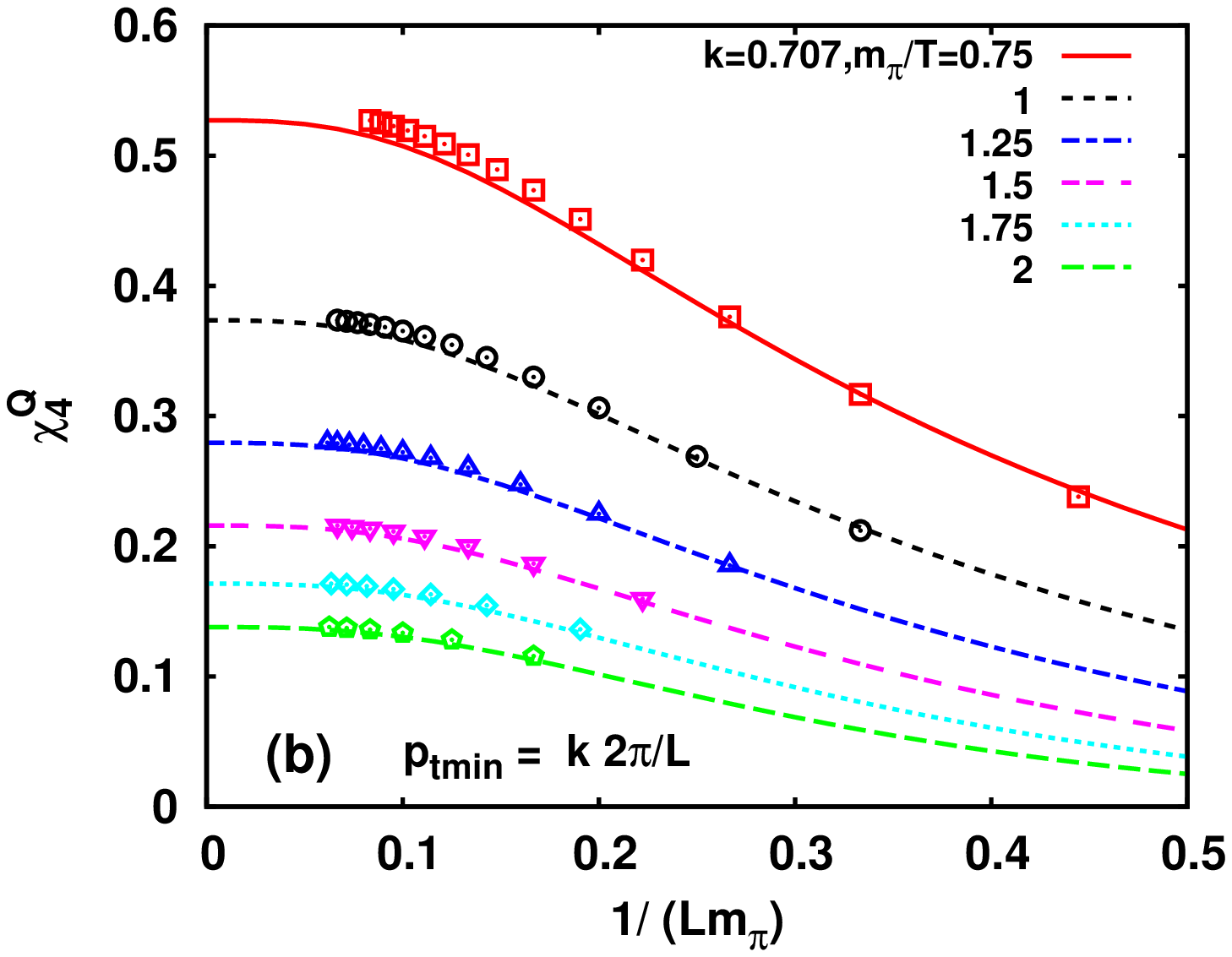}
 \caption{Quadratic [$\chi_2^Q$, (a)] and quartic [$\chi_4^Q$, (b)] cumulants of
 electric charge fluctuations in a non-interacting pion gas calculated
 in a finite box of size $L^3$  for different  $m_\pi/T$ (symbols). Results are
compared to calculations in an infinite box but with a non-zero transverse
momentum cut-off $p_{t_\text{min}}=2\pi k/L$ with $k=0.707$ (lines).
}
 \label{fig:volume}
\end{figure}

In Fig.~\ref{fig:volume} we show a calculation of
quadratic and quartic charge fluctuations of a pion gas in a finite
volume. Indeed, there is a clear correspondence between finite volume and
momentum cut-off effects in such a non-interacting gas. As can be seen
from Fig.~\ref{fig:volume},  all quadratic and quartic charge fluctuations
in a volume of size $L^3$ can be reproduced by a calculation of charge
fluctuations in an infinite box,  but with a transverse momentum cut-off
\begin{equation}
p_{t_\text{min}}=2\pi k/L \;\; {\rm with}\;\;  k=0.707 \; .
\label{correspondence}
\end{equation}
The proportionality factor $k$ is found to be rather  weakly depend on the pseudorapidity
acceptance. Indeed,
for $|\Delta \eta| = 0.35$ one gets  $k=0.707$, while for  $|\Delta \eta| = 5$
 one finds
$k\simeq 0.5$. 

In a non-interacting pion gas the scale for all dimensionfull quantities
is set by the temperature. Consequently, the size of a finite volume or a non-zero
momentum becomes meaningful in units of $T$. We thus may
interpret the above results in terms of  $LT$, which is
conventionally used in lattice field theory. In particular, lattice
QCD calculations are performed on lattices of size
$L^3T^{-1}\equiv (N_\sigma a)^3 N_\tau a$ and the continuum limit
($N_\sigma \rightarrow \infty, N_\tau\rightarrow \infty$)
is taken for fixed aspect ratio $N_\sigma/N_\tau\equiv LT$.
Typically $LT\simeq 4$
in such calculations. We may compare this to the momentum cut-off
$p_{t_\text{min}}\simeq (0.2-0.3)$~GeV currently used in heavy ion
experiments. At a freeze-out temperature $T\simeq 160$~MeV these
momentum cut-offs expressed in units of temperature are,
$p_{t_\text{min}}/T\simeq (1.25-1.875)$. Using Eq.~\eqref{correspondence},
which relates $p_{t_\text{min}}/T$ to a box size $LT$, we find that
these cut-offs correspond to aspect ratios, $LT\simeq 2.4-3.6$.
Thus, lattice calculations performed with an aspect ratio $LT=4$ seem
to be appropriate to characterize the thermodynamics probed with
transverse momentum cut-offs similar to those used by the STAR Collaboration.

\section{Concluding Remarks}
We have discussed the influence of kinematic cuts in transverse
momentum and pseudo-rapidity on  electric charge fluctuations in a static
and expanding medium.

Our basic arguments were given by
considering  a very transparent example of
an ideal   gas of charged pions with quantum statistics. Such a system
can be viewed  as a multi-component gas  of Boltzmann  particles with
different charges,  masses, and degeneracy factors,  thus  it contains  all relevant
properties to study charge  fluctuations,  with a  transparent
implementation   of different kinematic cuts.

We demonstrate,  that  a non-vanishing lower $p_t$ cut in the
net-electric charge fluctuations,  can substantially reduce  the second
and fourth order cumulants of net electric charge fluctuations, as well
as,  their ratio quantified by  the  normalized kurtosis
$\kappa_Q\sigma_Q^2$. These conclusions are valid  in a static and
expanding medium. The effect of pseudo-rapidity cuts was shown to be
rather negligible. Consequently, the observed  differences between STAR
and PHENIX data  on  cumulants of  electric charge fluctuations
could be traced back to their different  $p_t$  acceptance.

The experimental data on the ratio of mean and variance,
$M_Q/\sigma_Q^2$, and normalized kurtosis
$\kappa_Q\sigma_Q^2$,  were analyzed   in a hadron resonance gas.
It was shown, that in heavy ion collisions, the
influence of a non-zero transverse momentum cut-off, $p_{t_\text{min}}$, on $M_Q/\sigma_Q^2$
is almost independent of the collision energy, and that it increases,
with increasing  $p_{t_\text{min}}$.
Thus, the larger $p_{t_\text{min}}$  used by PHENIX,  suggests that their  $M_Q/\sigma_Q^2$
should be  larger than the corresponding
STAR values.  We have   shown,   that  indeed, by   rescaling the STAR and PHENIX data to vanishing transverse
momentum,  makes these data compatible within the still large systematic errors of the PHENIX data.
Furthermore, it was argued that  the reduction of $\kappa_Q\sigma_Q^2$
due to the shift of $p_{t_\text{min}}$   will lead to differences
between  PHENIX and STAR data of ${\cal O}(6\%)$,  which currently are buried under  large
statistical and systematic errors.

Having in mind a possible comparison between experimental data on
fluctuations of conserved charges  and lattice QCD results,  we have
discussed the relation between momentum cut-off and finite volume
effects. We have shown, in the context of an ideal pion gas,  that
cumulants of electric charge fluctuations  in a finite  box  of size
$L^3$,  can be reproduced by a calculation of charge
fluctuations in an infinite box,  but with a transverse momentum cut-off
$p_{t_\text{min}}=2\pi k/L$,   where  $k$ is a constant.

Finally, considering typical
box sizes used in lattice QCD calculations,  and the lower transverse
momentum cut-off scale in the  experimental data on electric charge
fluctuations, we have argued, that their direct comparison at a chiral
crossover temperature,  might be justified.

\acknowledgments
K.R.   acknowledges  stimulating discussions with Peter Braun-Munzinger,
Volker Koch  and Nu Xu. We also acknowledge fruitful  discussions with
Bengt Friman, Swagato Mukherjee and Chihiro Sasaki.
K.M. acknowledges the support of the  GSI/EMMI group  where a part of this work was completed.
This work was supported by the Grants-in-Aid for Scientific Research on
Innovative Areas from MEXT (No. 24105008), the Polish
Science Center (NCN) under Maestro grant 2013/10/A/ST2/00106,
the U.S. Department of Energy under Contract No. DE-SC0012704 and DE-FG02- 05ER41367, 
as well as, the Bundesministerium f\"ur Bildung und Forschung (BMBF)
under grant no. 05P15PBCAA.

\end{document}